# Chemical Pressure Effect on the Magnetic Order of the La$_{1.4}$Sr$_{1.6}$Mn$_2$O$_7$ Bilayered Manganite

Dr. Lorenzo Malavasi,*[a] Dr. Maria Cristina Mozzati[b], Dr. Clemens Ritter[c],

Prof. Carlo Bruno Azzoni[b], and Prof. Giorgio Flor[a]

[a] *Department of Physical Chemistry "M. Rolla", Viale Taramelli 16, Pavia, Italy.*
*Fax: +39 0382 987757; Tel: +39 0382 987757;*
*E-mail: lorenzo.malavasi@unipv.it*
[b] *Department of Physics "A. Volta", Via Bassi 6, Pavia, Italy.*
[c] *Institute Laue-Langevin, Boite Postale 156, F-38042, Grenoble, France.*

In recent years, there has been an increasing interest in studying the role of pressure, both external and internal (chemical), on the physical properties of different materials with the aim of shedding light on the basic properties of magnetic and other functional complex compounds. Significant efforts have been directed towards strongly correlated and layered materials such as manganites. Within this class most of the available theoretical and experimental work has, until now, been focused on the 3D perovskite structures, that is the $n = \infty$ end-member of the $A_{n+1}B_nO_{3n+1}$ Ruddlesden-Popper family, in which $n$ 2D layers of $BO_6$ corner-sharing octahedra are joined along the stacking direction and separated by rock-salt AO layers.

The optimally-doped $n = 2$ members of this family ($La_{2-2x}B_{1+2x}Mn_2O_7$ where B = Ca or Sr) behave analogously to the $n = \infty$ manganites in the sense that they undergo an insulating- to metallic-like state (I-M) transition coupled to a ferromagnetic transition at temperatures around 120-140 K; besides, they present a large magnetoresistance in this temperature range [1]. However, the change in dimensionality and resulting pronounced cation dependence of the electronic properties can produce physical properties which contrast strongly with the perovskite systems [2-6].

In this Communication we report about the effect induced by chemical pressure on the magnetic order within the $La_{1.4}Sr_{1.6}Mn_2O_7$ manganite. In particular, we prepared and studied, by means of neutron diffraction, the pure $La_{1.4}Sr_{1.6}Mn_2O_7$ sample and another sample in which about 25% of the Sr was replaced by the smaller Ca ($La_{1.4}Sr_{1.2}Ca_{0.4}Mn_2O_7$) while keeping the average Mn valence state constant.

Samples have been prepared by solid state reaction starting from proper amounts of $La_2O_3$, $Mn_2O_3$, $CaCO_3$ and $SrCO_3$ (Aldrich >99.99%). Pellets were prepared from the thoroughly mixed powders and allowed to react first at 1273 K for 72 hours and afterwards at 1573 K for 72 hours. During the thermal treatments pellets were re-ground and re-pelletized at least three times. In order to control the Mn valence state the final thermal treatment was carried out in pure argon.

Neutron diffraction data were collected on the D1A instrument at the ILL Facility in Grenoble with a wavelength of 1.39 Å at room temperature and of 1.91 Å at 137, 94, 52 and 2 K. Neutron data were refined by means of the FULLPROFILE software [7]. We stress that an impurity phase of perovskite structure has been detected and its amount estimated to be around 10%. The presence of this kind of spurious phase is very often encountered when working with double layered manganites.

Figure 1a reports the neutron diffraction patterns for $La_{1.4}Sr_{1.6}Mn_2O_7$ collected at four different temperatures while Figure 1b presents the same kind of measurements for the $La_{1.4}Sr_{1.2}Ca_{0.4}Mn_2O_7$ compound. Up to now we just measured the samples for a limited number of temperatures. However, even though the full development of the magnetic structure can not be followed through these $T$ values, we remark that they cover the most significant $T$-range regarding the magnetic transitions which are placed at around 100 K and 50 K for the pure and Ca-doped samples, respectively.

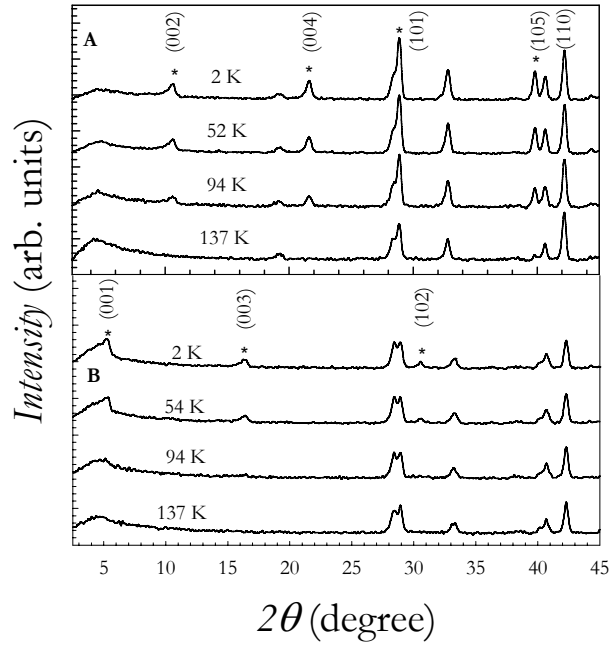

**Fig. 1** Neutron patter for $La_{1.4}Sr_{1.6}Mn_2O_7$ (a) and $La_{1.4}Sr_{1.2}Ca_{0.4}Mn_2O_7$ (b) at four different temperatures. Asterisks mark the magnetic peaks whose index are reported in parentheses.

Figure 1a, which refers to $La_{1.4}Sr_{1.6}Mn_2O_7$, shows the intensity increase of some nuclear peaks (marked with an asterisk in the Figure) starting from 94 K due to evolution of the magnetic contribution to the intensity. These peaks can be indexed considering a magnetic cell where all the Mn spins are aligned in the same direction within the *a-b* planes of the single bilayer and all the successive bilayers are aligned in the same direction thus giving origin to a ferromagnetic (FM) long-range order.

Figure 1b shows that for the $La_{1.4}Sr_{1.2}Ca_{0.4}Mn_2O_7$ sample only the nuclear peaks are visible at 137 and 94 K while, starting from the 54 K pattern, the magnetic peaks become evident. These peaks can be indexed according to an anti-ferromagnetic (AFM) spin arrangement of the A-type where each layer of the bilayer is FM coupled but the order within the bilayer is AFM. No traces of FM long-range order is present besides the antiferromagnetism.

Figure 2 shows the magnetic moment of the Mn ions for the two samples (in Bohr magnetons units) as calculated from the Rietveld refinements of the magnetic structure, as a function of temperature. For the $La_{1.4}Sr_{1.6}Mn_2O_7$ sample the value of the magnetic moment at 2 K is not far from the expected value of 3.70 $\mu_B$, while for the $La_{1.4}Sr_{1.2}Ca_{0.4}Mn_2O_7$ sample it is lower, thus suggesting that some spin-canting may occur. In fact, a possible rotation along the *c*-axis of the AFM structure may not be ruled out even though it would be very small since no traces of long-range FM peaks in the $La_{1.4}Sr_{1.2}Ca_{0.4}Mn_2O_7$ pattern have been found. Also the possibility of a canting with respect to the parallel alignment of the spin in the two layers composing the bilayers would give origin to FM peaks. This relatively small moment might also originate from some spin disorder connected to the Ca-doping which will lower the value of the net magnetic moment. However, the present data does not allow to get further insight into this last aspect that will be analysed by collecting more data especially at low angles (<5°).

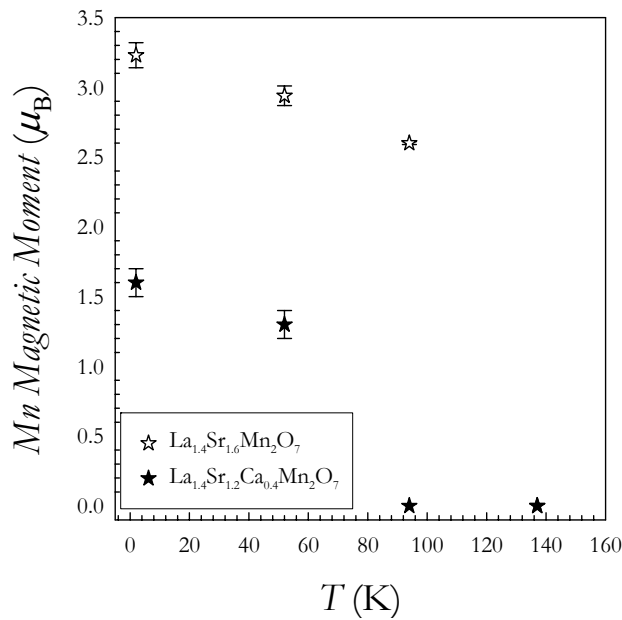

**Fig. 2** Magnetic moment as derived from the Rietveld refinements of the magnetic structures for $La_{1.4}Sr_{1.6}Mn_2O_7$ (empty stars) and $La_{1.4}Sr_{1.2}Ca_{0.4}Mn_2O_7$ (full stars) as a function of temperature.

The striking feature of our finding is the marked difference in the magnetic behaviour between two samples presenting the same amount of hole doping. The presence of the same Mn valence state was confirmed through extensive x-ray absorption spectroscopy (XAS) analysis at the Mn-K edge (shown as Figure 3 in the electronic supplementary information, ESI).

An analogous behaviour in the evolution of the magnetic structure in bilayered manganites, *i.e.* from FM to A-AFM, was found as a function of divalent (Sr) dopant concentration. In particular, it was shown that going from $x \approx 0.30$ to $x \approx 0.50$ in $La_{2-2x}Sr_{1+2x}Mn_2O_7$ leads to the evolution of the magnetic coupling within the bilayer from FM with different spin direction to layered AFM [2]. Such an evolution in the magnetic structure strongly depends on the doping level and in turn on the hole doping, but also on the change in the orbital state of the $e_g$ electrons. Indeed, Kimura *et al*. [8], have shown that as the Sr-doping increases, a significant reduction of the *c* lattice parameter and a nearly constant or slight increase in the *a*(*b*) parameter occurs; in turn, they observed a steep shrink in the Mn-O(2) bond-length along the *c*-axis. This was then linked to the progressive shift of electronic density from the $3z^2$-$r^2$ to the $x^2$-$y^2$ $e_g$ band as the *x*-value increases. Let us remember that in these double layered manganites three kinds of Mn-O bonds exist: two apical Mn-O bonds, labelled as Mn-O(2) and Mn(O)3, with different lengths, and four equal equatorial Mn-O(1) bonds (see Figure 4, ESI, for details on the atoms labelling).

In this communication we point out that a trend similar to that found as a function of *x* is also observed by applying a chemical pressure in the $La_{2-2x}Sr_{1+2x}Mn_2O_7$ system for $x \approx 0.30$. Table 1 reports some structural parameter derived from the neutron diffraction refinement of the room temperature data. A more complete study regarding these data will be published later.

**Table 1** Structural parameters derived from the neutron diffraction pattern at room temperature and for λ = 1.39 Å for the two samples.

|  | $La_{1.4}Sr_{1.6}Mn_2O_7$ | $La_{1.4}Sr_{1.2}Ca_{0.4}Mn_2O_7$ |
|---|---|---|
| $a$ (Å) | 3.8719(1) | 3.8710(2) |
| $c$ (Å) | 20.2116(8) | 19.962(1) |
| $V$ (Å$^3$) | 303.01(2) | 299.13(3) |
| **Mn-O(1)** (Å) | 1.902(1) | 1.906(1) |
| **Mn-O(2)** (Å) | 2.084(1) | 2.044(1) |
| **Mn-O(3)** (Å) | 1.936(1) | 1.935(1) |
| $R_{wp}$ | 6.02 | 7.11 |

As can be appreciated, the Mn-O(2) bond length is the most sensitive when replacing Sr with Ca, moving from 2.084(1) to 2.044(1), while the other two bonds are less sensitive to the doping (see Table 1 for the values).

These results suggest that the Ca-doping, at fixed hole concentration, induces a contraction of the lattice which changes the relative stability of the $e_g$ orbitals and consequently of the magnetic structure, which passes from long-range FM to long-range A-type AFM.

To the best of our knowledge this is the first report of this kind of behaviour in bilayered manganites for an optimal doping, *i.e.* $x \approx 0.30$. The only other available work in which substitutional effects on the magnetic properties of a bilayered manganites have been considered is due to Akimoto *et al*., [9]. In this paper, however, they explored the $x \approx 0.40$ concentration, which is very close to the boundary of the AFM order [2]. Finally, we stress that our data are in agreement with the work of Kimura and co-workers [10], in which they studied the charge-transport and magnetic properties of $La_{1.4}Sr_{1.6}Mn_2O_7$ as a function of applied hydrostatic pressure. They showed that the applied pressure influenced the charge and spin coupling through the stabilization of the $3d\ x^2-y^2$ orbital.

To conclude, in this Communication we have reported the stabilization of the A-type AFM long-range order for the $La_{1.4}Sr_{1.6}Mn_2O_7$ bilayered manganites induced by the partial replacement of the Sr with the smaller Ca, keeping constant the hole doping. This is in turn connected to the change in the orbital character of the $e_g$ electrons as a function of Ca-doping.

**SUPPLEMENTARY INFORMATION**

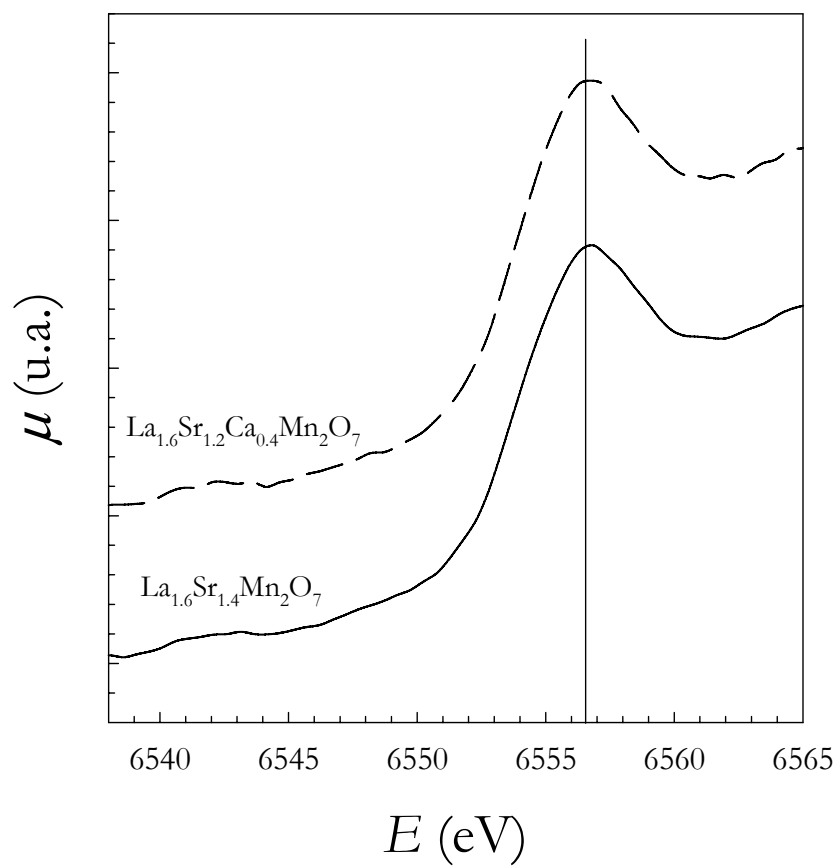

**Figure 3** – Mn-K edge for the two samples considered in the paper. Vertical line passes through the maximum of the spectra.

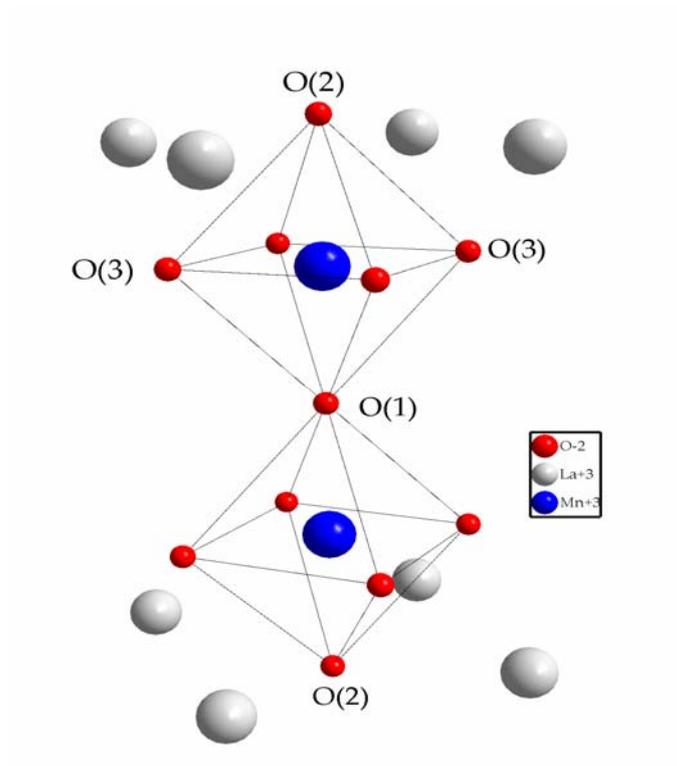

**Figure 4 -** Schematic representation of the Mn-O bonds type present in the bilayered manganites.